# Multigrid Monte Carlo Algorithms for $SU(2)$ Lattice Gauge Theory: Two versus Four Dimensions

**Martin Grabenstein**[1] **and Klaus Pinn**[2]

[1] Department of Applied Mathematics and Computer Science,
The Weizmann Institute of Science, Rehovot 76100, Israel,
internet i0̸2gra@dsyibm.desy.de

[2] Institut für Theoretische Physik I, Universität Münster,
Wilhelm-Klemm-Str. 9, 48149 Münster, Germany,
internet pinn@yukawa.uni-muenster.de

### Abstract

We study a multigrid method for nonabelian lattice gauge theory, the time slice blocking, in two and four dimensions. For $SU(2)$ gauge fields in two dimensions, critical slowing down is almost completely eliminated by this method. This result is in accordance with theoretical arguments based on the analysis of the scale dependence of acceptance rates for nonlocal Metropolis updates. The generalization of the time slice blocking to $SU(2)$ in four dimensions is investigated analytically and by numerical simulations. Compared to two dimensions, the local disorder in the four dimensional gauge field leads to kinematical problems.

# 1 Introduction

Standard Monte Carlo algorithms for the simulation of Euclidian lattice field theory close to the continuum limit suffer from critical slowing down (CSD): The autocorrelation time $\tau$, the time to obtain a new, useful measurement in a computer simulation, diverges with the correlation length $\xi$ like $\tau \sim \xi^z$. The dynamical critical exponent $z$ for local algorithms is $z \approx 2$. This means that close to the continuum limit, where the cutoff (and the correlation length) is increased, there is an enormous increase of computer time to calculate physical observables to a given accuracy.

The development of fast Monte Carlo algorithms that can overcome CSD in numerical simulations of nonabelian lattice gauge theory in four dimensions is an open problem. In particular the nonabelian gauge groups $SU(2)$ and $SU(3)$ are of relevance for nonperturbative calculations in the standard model of elementary particles.

For the dynamical critical exponent of the local heat bath algorithm in $SU(3)$ only very preliminary estimates are available up to now [1], consistent with $z \approx 2$. The present state-of-the-art algorithm for nonabelian gauge fields in four dimensions is overrelaxation [2]. For this algorithm, first estimates for $z$ in $SU(2)$ lattice gauge theory yielded $z = 1.0(1)$ in physically small volumes [3].

Methods that have been developed for spin models in order to overcome CSD completely in the sense of $z \approx 0$ are nonlocal algorithms: stochastic cluster algorithms [4] and multigrid Monte Carlo techniques [5]. The status of nonlocal updating algorithms applied to continuous gauge fields is as follows:

A fast cluster algorithm was found for $3 + 1$-dimensional $SU(2)$ gauge theory on a $L^3 \times T$ lattice at finite temperature, but only in the special case $T = 1$ [6]. A cluster algorithm for $U(1)$ gauge theory in two dimensions based on the reduction of the gauge theory to a one dimensional $XY$ model was developed [7]. Apart from these special cases, no efficient cluster algorithm for continuous gauge groups has been found up to now.

Therefore the development of stochastic multigrid methods for pure gauge fields is of particular interest. Multigrid algorithms for $U(1)$ gauge models were introduced and studied in two and four dimensions [8, 9]. A different but related nonlocal updating scheme for abelian lattice gauge theory in four dimensions is the multiscale method [10]. However, since the phase transition of $U(1)$ in four dimensions is believed to be a weak first order transition, it is not straightforward to judge on the potential of these methods to reduce CSD.

An attempt to understand *why* multigrid Monte Carlo is successful in beating CSD for some models while it does not work as well for others, was made in [11, 12]: An analytic calculation and analysis of acceptance rates for nonlocal Metropolis updating was performed. The scale dependence of acceptance rates for interacting models was compared with the behavior in free field theory, where CSD is known to be eliminated by a multigrid algorithm. By this kinematical analysis one can predict whether a specific multigrid procedure will have the potential to overcome CSD in the simulation of a given model.

We will use this analysis as a guideline for the development of multigrid algorithms for nonabelian gauge fields. To gain experience, we first study the case of gauge group $SU(2)$ in two dimensions. We introduce a multigrid method for nonabelian gauge theory that treats different time slices independently: the time slice blocking algorithm. The theoretical analysis predicts that CSD can be eliminated by the time slice blocking. By numerical experiments on systems with lattice sizes up to $256^2$ we check whether this is indeed the case.

In a second step we generalize the time slice blocking to $SU(2)$ in four dimensions. Compared to the two dimensional case we have to face additional difficulties that are caused by the local disorder of the gauge fields. We attempt to estimate the kinematical behavior of the proposed algorithm in the weak coupling limit and study whether a reduction of CSD can be expected. In numerical simulations, the time slice blocking algorithm in $SU(2)$ in four dimensions is compared with a local heat bath algorithm.

This paper is organized as follows: Section 2 is a brief review of the analysis of multigrid algorithms. Most of the concepts that are needed for the treatment of nonabelian gauge fields are introduced in section 3 in the context of $SU(2)$ lattice gauge theory in two dimensions. A multigrid procedure, the time slice blocking, is introduced. In section 4, $SU(2)$ lattice gauge theory in two dimensions is simulated with the time slice blocking. The procedure is generalized from $SU(2)$ in two dimensions to $SU(2)$ in four dimensions in section 5. Section 6 reports on a multigrid Monte Carlo simulation of $SU(2)$ lattice gauge theory in four dimensions. A summary is given in section 7.

# 2 Kinematical Analysis of Multigrid Algorithms

## 2.1 Multigrid Algorithms for Spin Models

We consider models with partition functions

$$Z = \int \prod_{x \in \Lambda_0} d\phi_x \, \exp(-\mathcal{H}(\phi)) \tag{2.1}$$

on cubic $d$-dimensional lattices $\Lambda_0$. We shall use dimensionless spin variables. Nonlocal Monte Carlo updates are defined as follows: Divide the fundamental lattice $\Lambda_0$ in cubic blocks of size $l^d$ (e.g. $l = 2$). This defines a block lattice $\Lambda_1$. By iterating this procedure one gets a hierarchy of block lattices $\Lambda_0, \Lambda_1, \ldots, \Lambda_K$. We denote block lattice points in $\Lambda_k$ by $x'$. Block spins $\Phi_{x'}$ are defined on block lattices $\Lambda_k$. They are averages of the fundamental field $\phi$ over blocks of side length $L_B = l^k$:

$$\Phi_{x'} = L_B^{(d-2)/2} L_B^{-d} \sum_{x \in x'} \phi_x \,. \tag{2.2}$$

A nonlocal change of the configuration $\phi$ consists of a shift

$$\phi_x \to \phi_x + s \, \psi_x, \tag{2.3}$$

where $s$ is a real parameter. The shape of the nonlocal change is determined by the "coarse-to-fine interpolation kernel" $\psi$ that obeys the constraint

$$L_B^{-d} \sum_{x \in x'} \psi_x = L_B^{(2-d)/2} \delta_{x', x'_o} \,. \tag{2.4}$$

Note that the effect of (2.3) on $\Lambda_k$ is $\Phi_{x'} \to \Phi_{x'} + s$ for $x' = x'_o$, whereas $\Phi_{x'}$ remains unchanged on the other blocks. The simplest $\psi$ is a piecewise constant kernel: $\psi_x = L_B^{(2-d)/2}$ if $x \in x'_o$, and 0 else. One can also use smooth kernels that avoid large energy costs from the block boundaries.

The $s$-dependent Metropolis acceptance rate for such proposals is given by

$$\Omega(s) = \langle \min[1, \exp(-\Delta \mathcal{H})] \rangle, \tag{2.5}$$

where $\Delta \mathcal{H} = \mathcal{H}(\phi + s\psi) - \mathcal{H}(\phi)$.

## 2.2 Analysis of acceptance rates

The starting point of our acceptance analysis is the approximation formula [13, 11, 12]

$$\Omega(s) \approx \mathrm{erfc}\left(\tfrac{1}{2}\sqrt{h_1}\right). \tag{2.6}$$

Here, $h_1 = \langle \Delta \mathcal{H} \rangle$ denotes the average change in the fundamental Hamiltonian. Generally, the formula yields precise estimates that are confirmed by the acceptance rates directly measured in Monte Carlo simulations [11, 12]. We will use (2.6) to predict the acceptance rate $\Omega(s)$ for interacting models. Let us first discuss free massless field theory with action $\mathcal{H}(\phi) = \frac{1}{2}(\phi, -\Delta \phi)$. Here, we obtain the exact result

$$\Omega(s) = \mathrm{erfc}(\sqrt{\alpha/8}|s|), \tag{2.7}$$

with $\alpha = (\psi, -\Delta \psi)$. In $d$ dimensions one finds $\alpha = 2dL_B$ for piecewise constant kernels, and, for smooth kernels, $\alpha \to$ const if $L_B \gg 1$. (For a systematic study of different kernels see ref. [12].) As a consequence, in massless free field theory, to maintain a constant acceptance rate (of, say, 50 percent) the amplitudes $s$ have to be scaled down like $L_B^{-1/2}$ for piecewise constant kernels, whereas for smooth kernels the acceptance rates do not depend on the block size. At least for free field theory, the disadvantage of the piecewise constant kernels can be compensated for by using a W-cycle instead of a V-cycle. Smooth kernels can be used only in V-cycle algorithms.

The kinematical analysis of interacting models is a comparison of the scale dependence of acceptance rates for interacting models with the behavior in free field theory, where CSD is known to be eliminated by a multigrid algorithm.

Analyzing multigrid algorithms, two classes of models were found [11, 12]: For the first class, $s$ has to be rescaled like $L_B^{-1}$ for piecewise constant and for smooth kernels. Compared to free field theory, this is a dramatic decrease of acceptance when the blocks become large. It is therefore unlikely that any multigrid algorithm - based on nonlocal updates of the type discussed here - will be successful for such models.[1]

For the second class, the scale behavior of acceptance rates is as in free field theory. For almost all models of the second class, at least a substantial reduction of CSD could be achieved [15, 16, 8]. An exception is the 2d $XY$ model in the vortex phase. There $z \approx 1.4$ was found [17]. This shows that good acceptance rates alone are not sufficient to overcome CSD.

---
[1] A simple random walk argument suggests that a higher multigrid cycle could overcome this difficulty. At least in the Sine-Gordon model this is not the case [14].

All results of the analysis are consistent with the following rule: Sufficiently high acceptance rates for a complete elimination of CSD can only be expected if $h_1 = \langle \mathcal{H}(\phi+s\psi)-\mathcal{H}(\phi)\rangle$ contains no algorithmic "mass" term $\sim s^2 \sum_x \psi_x^2$. Formulated differently, an intuitive guideline for the development of new multigrid methods is [18]: *A piecewise constant update of a nonlocal domain should have energy costs proportional to the surface of the domain, but not energy costs proportional to the volume of the domain.*

We will use this heuristic criterion in the development of multigrid algorithms for gauge fields. The intuitive guideline can always be made precise by performing a quantitative acceptance analysis.

## 3 Multigrid algorithms for nonabelian lattice gauge fields in two dimensions

Let us introduce the notations for lattice gauge theory in $d$ dimensions. We consider partition functions
$$Z = \int \prod_{x,\mu} dU_{x,\mu} \, \exp(-\mathcal{H}(U)). \tag{3.1}$$

The link variables $U_{x,\mu}$ take values in the gauge group $U(1)$ or $SU(N)$, and $dU$ denotes the corresponding invariant Haar measure. The standard Wilson action $\mathcal{H}(U)$ is given by
$$\mathcal{H}(U) = \beta \sum_{\mathcal{P}} [1 - \tfrac{1}{N} \operatorname{Re} \operatorname{Tr} U_{\mathcal{P}}]. \tag{3.2}$$

The sum in (3.2) is over all plaquettes in the lattice. The $U_{\mathcal{P}}$ are path ordered products around plaquettes $\mathcal{P}$,
$$U_{\mathcal{P}} = U_{x,\mu} U_{x+\hat{\mu},\nu} U^*_{x+\hat{\nu},\mu} U^*_{x,\nu}. \tag{3.3}$$
$U^*$ denotes the hermitean conjugate (= inverse) of $U$.

### 3.1 Basic idea

A first multigrid Monte Carlo algorithm for abelian gauge fields was implemented and tested by Laursen, Smit and Vink for $U(1)$ lattice gauge theory in two dimensions [8]. It was inspired by a deterministic multigrid procedure for the minimization of a Hamiltonian of the form arising in lattice gauge theory [19]. We review their basic updating concept in the unigrid language.

Nonlocal updates are defined as illustrated in figure 1: One chooses a square block $x'_o$ of size $L_B^2$ and a direction $\tau$ with $\tau = 1$ or 2. During the update, $\tau$ will be kept fixed. All the link variables $U_{x,\tau}$ attached to sites $x$ inside the block $x'_o$ are proposed to be "rotated" simultaneously:
$$U_{x,\tau} \to R_x U_{x,\tau}, \tag{3.4}$$
where the "rotation" matrix $R_x$ is taken from the gauge group $U(1)$. In the reported work piecewise constant rotation matrices $R_x = R$ for $x \in x'_o$, $R_x = 1$ for $x \notin x'_o$ were used.

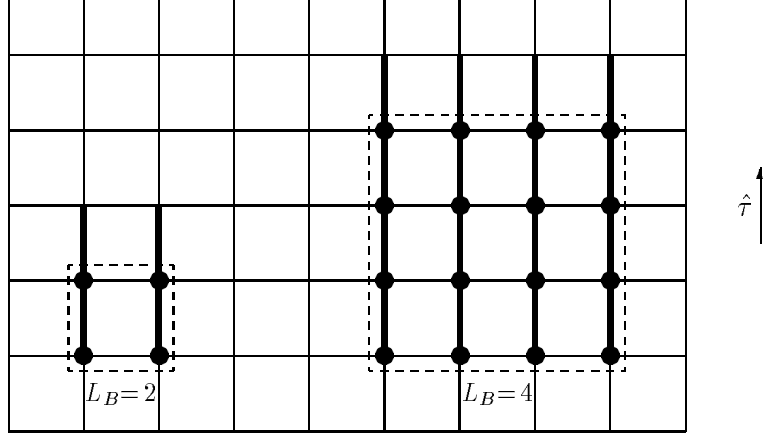

Figure 1: *Blocking of link variables for two dimensional $U(1)$ gauge fields*

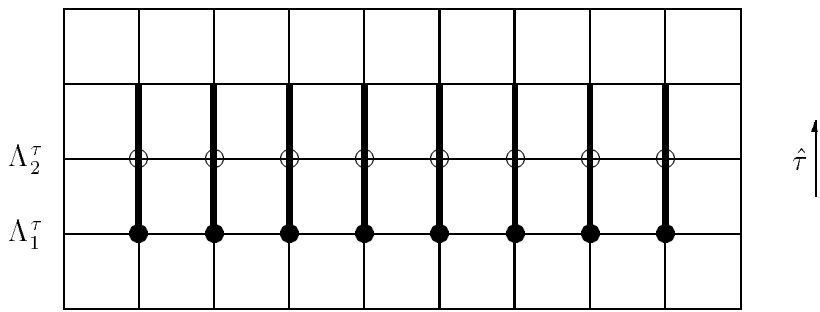

Figure 2: *Decoupling of time slices $\Lambda_1^\tau$ and $\Lambda_2^\tau$.*

## 3.2  Comments and possible modifications

Let us denote the time slice of lattice sites with $\tau$-component $t$ as $\Lambda_t^\tau = \{x \in \Lambda_0 \,|\, x_\tau = t\}$. Here the name "time" direction has no physical meaning. We use this word to label the fixed direction of link variables that are updated simultaneously. The time direction in an update algorithm will be changed periodically from $\tau = 1$ to $\tau = 2$. In the following, we will denote the time direction with $\tau$ and the spatial direction(s) different from $\tau$ with $\mu$.

In the unigrid picture a general feature of nonlocal updating schemes in lattice gauge theory is transparent: As long as we restrict the possible nonlocal changes in the configuration to link variables $U_{x,\tau}$ of a fixed time direction $\tau$ and keep all other variables $U_{x,\mu}$ with $\mu \neq \tau$ unchanged, adjacent time slices decouple. This is illustrated in figure 2.

Here, link variables $U_{x,\tau}$ pointing from sites in two different adjacent time slices $\Lambda_1^\tau$ and $\Lambda_2^\tau$ are shown. The point is that there are only plaquette terms in the Hamiltonian that contain link variables $U_{x,\tau}$ pointing from sites $x$ in the same time slice $\Lambda_t^\tau$, i.e. either in $\Lambda_1^\tau$ or $\Lambda_2^\tau$. This means that the link variables pointing from sites in $\Lambda_1^\tau$ and from sites in $\Lambda_2^\tau$ are independent as long as all other $U_{x,\mu}$ with $\mu \neq \tau$ are fixed. In particular, the rotation matrices $R_x$ need not to be smooth in the $\tau$-direction. From now on we choose the $R_x$ to be constant in that direction.

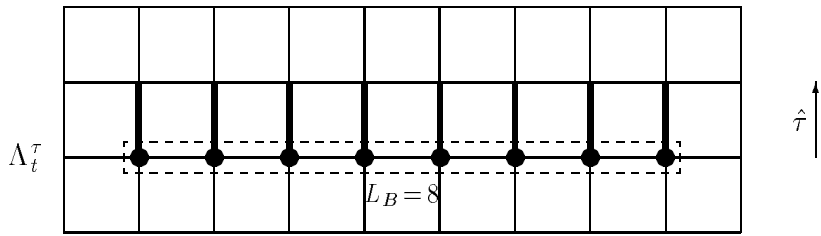

Figure 3: *Geometry of the time slice blocking: The marked link variables are updated simultaneously. The bottom of the block is indicated by a dashed line.*

Since updates of link variables in different time slices are statistically independent, two different nonlocal update schemes are possible: the time slice blocking and the square blocking. In the time slice blocking method updates of one dimensional blocks of size $L_B$ are performed on separate time slices $\Lambda_t^\tau$ in sequence. In the square blocking scheme as used by Laursen, Smit and Vink one builds $L_B \times L_B$ blocks out of "staples" of $L_B$ one dimensional blocks of size $L_B$ and performs the updates on this square block simultaneously. The analysis of the kinematics is the same for both schemes. For simplicity we are going to adopt the time slice blocking in the following.

An important point is that the decoupling of time slices is independent of the gauge group and carries over to higher dimensions. We are going to use this fact as a basic ingredient of nonlocal updating schemes for nonabelian gauge fields.

## 3.3 The nonabelian character of the gauge field

Our method for a nonlocal updating procedure for nonabelian $SU(2)$ gauge fields in two dimensions is based on the time slice blocking. It is illustrated in figure 3. We start the discussion with a naive generalization of the updates in the abelian case: Choose a one dimensional block $x'_o$ of size $L_B$ within a time slice $\Lambda_t^\tau$ and update all link variables $U_{x,\tau}$ pointing from this block in the $\tau$-direction,

$$U_{x,\tau} \to U'_{x,\tau} = R_x U_{x,\tau} \quad \text{for all} \ \ x \in x'_o \, , \qquad (3.5)$$

where the "rotation" matrices $R_x$ are in $SU(2)$. We parametrize them as

$$R_x(\vec{n}, s) = \cos(s\psi_x/2) + i \sin(s\psi_x/2)\, \vec{n} \cdot \vec{\sigma} \, , \qquad (3.6)$$

where $\vec{n}$ denotes a three dimensional real unit vector, and the $\sigma_i$ are Pauli matrices. $\vec{n}$ will be taken randomly from the three dimensional unit sphere, and $\psi$ will have support on the one dimensional block $x'_o$.

The simplest version is a piecewise constant "rotation", where $R_x$ is a rotation matrix $R$ independent of $x$. Let us examine how a plaquette in the interior of the block (as illustrated

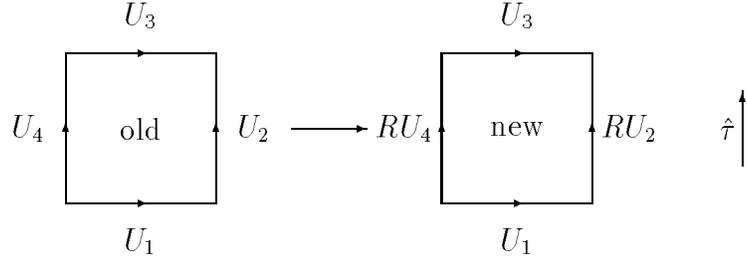

Figure 4: *Change of plaquette during naive nonlocal updating*

in figure 4) changes under this update:

$$P_{old} = \tfrac{1}{2}\text{Tr}(U_1 U_2 U_3^* U_4^*) \rightarrow$$

$$P_{new} = \tfrac{1}{2}\text{Tr}(U_1 R U_2 U_3^* (R U_4)^*) = \frac{1}{2}\text{Tr}(R^* U_1 R U_2 U_3^* U_4^*) \tag{3.7}$$

with the notations as in figure 4. Let denote the links

$$(x, x+\mu) \text{ for all } x, x+\mu \in x'_o \tag{3.8}$$

as the links in the bottom of the block, as marked in figure 3. In the abelian case the rotation matrix $R$ and the link variable in the bottom of the block $U_1$ would commute, and we would have $P_{old} = P_{new}$, i.e. the plaquettes in the interior of the block would remain unchanged.

But since our gauge group is nonabelian the resulting change of the plaquette by this naive generalization would lead to energy costs proportional to the volume of the block.

As formulated in the intuitive guideline at the end of section 2, energy costs proportional to the volume of the block lead to an algorithmic mass term that suppresses the amplitudes of the nonlocal moves on large scales. Therefore the naive generalization will suffer from CSD.

This does not come as a surprise. Due to the gauge invariance of the model, the link variables do not have a gauge invariant meaning. Therefore the rotation $R$ that is constant over the block for a given gauge can be arbitrarily rough and disordered after a gauge transformation. It is therefore natural to assume that the rotation matrices have to be chosen in a gauge covariant way.

## 3.4  Gauge covariant time slice blocking algorithm

We use the additional gauge degrees of freedom and generalize (3.5) to

$$U_{x,\tau} \rightarrow U'_{x,\tau} = R_x(g) U_{x,\tau} \text{ for all } x \in x'_o, \tag{3.9}$$

with $R_x(g) = g_x^* R_x g_x$ and $g$-matrices $g_x \in SU(2)$. In the abelian case we obtain nothing new, because $g_x$ and $R_x$ commute. In the nonabelian case a plaquette in the interior of the block changes under a piecewise constant rotation $R$ according to

$$P_{old} = \tfrac{1}{2}\text{Tr}(U_1 U_2 U_3^* U_4^*) \rightarrow P_{new} = \tfrac{1}{2}\text{Tr}(R^* U_1^g R U_2^g U_3^{g*} U_4^{g*}). \tag{3.10}$$

The gauge transformed link variables are given by $U^g_{x,\mu} = g_x U_{x,\mu} g^*_{x+\mu}$. If we choose the $g$-matrices in such a way that the link variable $U^g_1$ in the bottom of the block is equal to unity, it will commute with an arbitrary rotation matrix $R$. Then the plaquettes in the interior of the block stay unchanged as in the abelian case. (The bottom of the block is indicated in figure 3.)

From this discussion we are led to the following gauge condition: Choose the $g$-matrices in the bottom of the one dimensional block $x'_o$ such that

$$U^g_{x,\mu} = g_x U_{x,\mu} g^*_{x+\hat{\mu}} = 1 \quad \text{for all} \quad (x, x+\hat{\mu}) \in x'_o \,. \tag{3.11}$$

In other words: All gauge transformed link variables in the one dimensional bottom of the block should be equal to unity. We denote this gauge condition as block axial gauge. Note that we still have the freedom of a constant gauge transformation within each block $x'$,

$$g_x \to h_{x'} g_x \quad \text{for all} \quad x \in x' \,. \tag{3.12}$$

Although we use the term "gauge condition" here, we do not intend to perform the gauge transformation $g$. We use the concept of gauging only to define covariant rotations $R_x(g) = g^*_x R_x g_x$.

The gauge transformation properties of these updates are as follows: If we apply an arbitrary gauge transformation $h$ to the gauge field $U$

$$U_{x,\mu} \to U^h_{x,\mu} = h_x U_{x,\mu} h^*_{x+\hat{\mu}} \tag{3.13}$$

the $g$-matrices transform like

$$g_x \to g^h_x = g_x h^*_x \tag{3.14}$$

and the covariant rotation matrix $R_x(g) = g^*_x R_x g_x$ transforms according to the adjoint representation:

$$R_x(g) \to (R_x(g))^h = h_x R_x(g) h^*_x \,. \tag{3.15}$$

As a consequence, if we apply the updates to the gauge transformed configuration $U^h$

$$U^h_{x,\tau} \to (U^h_{x,\tau})' = (R_x(g))^h U^h_{x,\tau} = h_x R_x(g) h^*_x h_x U_{x,\tau} h^*_{x+\hat{\tau}}$$

$$= h_x R_x(g) U_{x,\tau} h^*_{x+\hat{\tau}} = h_x U'_{x,\tau} h^*_{x+\hat{\tau}} = (U'_{x,\tau})^h \,, \tag{3.16}$$

the updating commutes with a gauge transformation $h$ and is therefore gauge covariant.

Let us summarize the steps of the time slice blocking scheme for $SU(2)$ in the unigrid language:

1. Choose a block $x'_o$ of size $L_B$ that is contained in the slice $\Lambda^\tau_t$. All link variables $U_{x,\tau}$ pointing from sites $x$ inside the block in $\tau$-direction will be moved simultaneously.

2. Find the gauge transformation $g$ defined by the block axial gauge condition (3.11) such that $U^g_{x,\mu} = 1$ for all link variables in the bottom of the block.

3. Propose new link variables $U'_{x,\tau}$ by

$$U_{x,\tau} \to U'_{x,\tau} = R_x(g) U_{x,\tau}, \qquad (3.17)$$

with $R_x(g) = g_x^* R_x g_x$ and

$$R_x(\vec{n}, s) = \cos(s\psi_x/2) + i \sin(s\psi_x/2)\, \vec{n}\cdot\vec{\sigma}. \qquad (3.18)$$

$s$ is a uniformly distributed random number from the interval $[-\varepsilon, \varepsilon]$, $\vec{n}$ is a vector selected randomly from the three dimensional unit sphere, and $\psi$ is a one dimensional kernel.

4. Calculate the associated change of the Hamiltonian $\Delta\mathcal{H}$ and accept the proposed link variables with probability $\min[1, \exp(-\Delta\mathcal{H})]$.

The detailed balance condition is fulfilled by this updating scheme: For the naive version with $g = 1$ it is straightforward to show that the detailed balance condition holds, since the rotation matrices $R_x$ are chosen according to a probability distribution which is symmetric around unity.

If we now take $g$ according to some gauge condition, we have to be careful that we get the same $g$ before and after the move $U_{x,\tau} \to U'_{x,\tau}$ is performed. Otherwise this move would not be reversible. In other words: The gauge condition yielding $g$ must not depend on $U_{x,\tau}$. This is indeed the case, since only link variables $U_{x,\mu}$ with $\mu \neq \tau$ enter in the block axial gauge condition.

The details of an implementation and simulation of the covariant time slice blocking algorithm for $SU(2)$ gauge fields will be described in section 4.

## 3.5 Acceptance analysis of the proposal

The energy change associated with the update proposal (3.9) is

$$\Delta\mathcal{H} = -\frac{\beta}{2} \sum_{\mathcal{P}} \mathrm{Tr}(U'_{\mathcal{P}} - U_{\mathcal{P}}) = -\frac{\beta}{2} \sum_{\substack{x \in \Lambda_t^\tau \\ \mu \neq \tau}} \mathrm{Tr}\{(R_x(g)^* U_{x,\mu} R(g)_{x+\hat{\mu}} - U_{x,\mu}) H^*_{x,\mu}\}, \qquad (3.19)$$

where $H^*_{x,\mu} = U_{x+\hat{\mu},\tau} U^*_{x+\hat{\tau},\mu} U^*_{x,\tau}$, and $\psi$ stands for a one dimensional interpolation kernel. The relevant quantity for the acceptance rates is $h_1 = \langle \Delta\mathcal{H} \rangle$. If we assume that the $g$-matrices are chosen according to the block axial gauge condition and that $\psi$ vanishes outside the block $x'_o$ we find

$$h_1 = \beta P \sum_{\substack{x \in \Lambda_t^\tau \\ \mu \neq \tau}} [1 - \cos(s(\psi_x - \psi_{x+\hat{\mu}})/2)]$$

$$= \frac{s^2}{8} \beta P \sum_{\substack{x \in \Lambda_t^\tau \\ \mu \neq \tau}} (\psi_x - \psi_{x+\hat{\mu}})^2 + O(s^4) = \frac{s^2}{8} \beta P \alpha_1 + O(s^4) \qquad (3.20)$$

with $P = \langle \frac{1}{2}\mathrm{Tr}U_{\mathcal{P}} \rangle$ and $\alpha_1 = (\psi, -\Delta\psi)$. Remember that for the time slice blocking in two dimensions $\psi$ is a one dimensional kernel.

Thus the kinematical behavior of this method is the same as that of the massless Gaussian model in one dimension. Since we do not expect any additional topological problems to occur for $SU(2)$ in two dimensions, we expect a successful acceleration of the simulation by the proposed time slice blocking algorithm. This prediction will be verified in in section 4.

# 4 Multigrid Monte Carlo simulation of $SU(2)$ lattice gauge fields in two dimensions

## 4.1 Implementation of the time slice blocking

For a concrete implementation of the time slice blocking method as introduced in section 3 we choose a unigrid algorithm with piecewise linear interpolation and a V-cycle.

Although the analysis of nonlocal updates in section 3 was performed in terms of a Metropolis version, we are going to use a heat bath version of the nonlocal moves in our simulation. We think that the change of the update method from Metropolis to heat bath will not affect the dynamical critical behavior in a substantial way. The advantage of the heat bath updating is that there are no tuneable parameters such as the Metropolis step size $\varepsilon(L_B)$.

A heat bath implementation of nonlocal time slice blocking updates is possible if we use one dimensional piecewise linear interpolation and update in $U(1)$ subgroups of $SU(2)$. This will be described in the following.

Compared to section 3, we formulate the piecewise linear interpolation in a different language: Assume that the $g$-matrices defined according to the block axial gauge condition (3.11) have been applied as a gauge transformation in the bottom of the block $x'_o$. Then the gauged link variables in the bottom of the block are equal to unity. Now a piecewise linear block update is formulated by multiplying the $L_B$ gauged link variables $U^g_{x,\tau}$ pointing in $\tau$-direction from left to right by the $SU(2)$-matrices $R, R^2, R^3, \ldots R^{L_B/2}, R^{L_B/2}, R^{L_B/2-1}, \ldots, R$. $R$ is given by

$$R(\vec{n}, \theta) = \cos(\theta) + i \sin(\theta)\, \vec{n} \cdot \vec{\sigma}\,, \tag{4.1}$$

where the randomly chosen three-dimensional vector $\vec{n}$ specifies the direction of a $U(1)$ subgroup in $SU(2)$. All changes of plaquettes that are generated by this update can be written in the form

$$\mathrm{Tr}(U'_\mathcal{P}) = \mathrm{Tr}(RV) = \mathrm{Tr}(V)\cos(\theta) + \mathrm{Tr}(i\vec{n}\cdot\vec{\sigma}V)\sin(\theta)\,, \tag{4.2}$$

with the $SU(2)$ matrix $V = U^g_\mathcal{P}$ or $V = U^{g*}_\mathcal{P}$. By summing over all changed plaquettes (4.2) we obtain an overall change in the Hamiltonian of the form

$$\mathcal{H}(U') = a\cos(\theta) + b\sin(\theta) + const\,, \tag{4.3}$$

with real constants $a$ and $b$. To generate $U(1)$ random numbers distributed according to the distribution

$$\mathrm{dprob}(\theta) \propto \mathrm{e}^{-a\cos(\theta) - b\sin(\theta)} d\theta \tag{4.4}$$

we use the fast vectorizable method of Hattori and Nakajima [20].

The sequence of updates is organized as follows: We start with a V-cycle of time slice blocking updates on the links $U_{x,\tau}$ pointing in the $\tau = 1$ direction. The largest block size is $L/2$ on a $L \times L$ lattice. This means that the sequence of updated block sizes is $L_B = 2, 4, \ldots L/2, L/2, L/4 \ldots 2$. When all time slices have been updated by a V-cycle, we perform a sweep of local $SU(2)$ heat bath updates through all links on the lattice. Here we use the "incomplete" Kennedy-Pendleton [21] heat bath algorithm with one trial per link. This means that in a sweep through the lattice not all links but only a very high percentage of them are updated. The advantage is that scalar operations on a vector computer are avoided [22]. Then we do a V-cycle on all links pointing in the $\tau = 2$ direction and again a local heat bath sweep. This sequence is repeated periodically.

Measurements are performed after each local heat bath sweep. To avoid effects from fixed block boundaries, we use stochastically overlapping blocks [16] by applying a random translation before each V-cycle.

In order to save computer time we use a slight modification of the gauge condition. Recall the block axial gauge (3.11): Take the $g$-matrices in the bottom of a block $x'_o$ such that

$$U^g_{x,\mu} = g_x U_{x,\mu} g^*_{x+\hat{\mu}} = 1 \quad \text{for all} \quad (x, x+\hat{\mu}) \in x'_o \,. \tag{4.5}$$

We modify it to the axial gauge: Take the $g$-matrices in the bottom of a time slice $\Lambda^\tau_t$ such that

$$U^g_{x,\mu} = g_x U_{x,\mu} g^*_{x+\hat{\mu}} = 1 \quad \text{for} \quad (x, x+\hat{\mu}) \in \Lambda^\tau_t, \, x_\mu = 1, \ldots L - 1 \,. \tag{4.6}$$

In other words: all spatial links but the last link are gauged to one within the bottom of a time slice.

The computational advantage of this slight change in the gauge condition is that for the block axial gauge the $g$-matrices have to be calculated for each block lattice individually. For the axial gauge the $g$-matrices are the same for all block lattices and they do not change during the updating on different block lattices with different $L_B$. Therefore we have to calculate them only once before performing the entire V-cycle. The gauge covariance properties of the update are not affected by this modification.

## 4.2 Simulation and results

The observables measured are square Wilson loops $W(I, I) = \langle \frac{1}{2} \text{Tr}(U(C_{I,I}) \rangle$, where $U(C_{I,I})$ is the parallel transporter around a rectangular Wilson loop $C_{I,I}$ of size $I \times I$. On an $L \times L$-lattice we measure $W(1,1), W(2,2), W(4,4), W(8,8), \ldots, W(L/2, L/2)$. Another important class of quantities is built up from timelike Polyakov loops. A Polyakov loop at the one dimensional spatial point $r$ is defined by $P_r = \frac{1}{2} \text{Tr} \prod_{t=1}^{L} U_{(r,t),2}$ We measure the lattice averaged Polyakov loop $\bar{P} = \langle 1/L \sum_{r=1}^{L} P_r \rangle$ and the lattice averaged Polyakov loop squared $\bar{P}^2 = \langle (1/L \sum_{r=1}^{L} P_r)^2 \rangle$

In order to investigate the dynamical critical behavior of the multigrid algorithm, we simulate a sequence of lattices with fixed physical size $L \approx 10\xi$, where the correlation length $\xi$ is related to the string tension by $\kappa$ by $\xi = 1/\sqrt{\kappa}$. Then we have to use lattice sizes and $\beta$ values such that $L^2/\beta$ is constant. If we choose this large ratio of $L/\xi$, finite size effects are negligible. The detailed run parameters are given in table 1. The quoted correlation length is calculated by the exact solution [23] in the infinite volume limit. We started our runs from ordered configurations with all link variables set equal to unity. After equilibration, measurements were taken after each local heat bath sweep through the lattice.

Table 1: Run parameters for the multigrid Monte Carlo simulation of two dimensional $SU(2)$ lattice gauge theory.

| $\beta$ | 4 | 16 | 64 | 256 | 1024 |
|---|---|---|---|---|---|
| $L$ | 16 | 32 | 64 | 128 | 256 |
| $\xi$ | 1.55 | 3.22 | 6.51 | 13.05 | 26.12 |

The static results of the simulation are given in table 2. All our results for the Wilson loops are consistent with the exact solution [23] in the infinite volume limit. Here only results for $W(\frac{L}{16}, \frac{L}{16})$ and $W(\frac{L}{8}, \frac{L}{8})$ are quoted, which are the loops of about the size of a correlation length squared. We observed that these loop sizes have the largest autocorrelation times among the Wilson loops.

In general we found a very fast decorrelation of subsequent configurations in the Markov chain. Typically the autocorrelation function $\rho(t)$ dropped to zero within errors after $3-5$ measurements. Therefore it was impossible to look for an exponential regime in the decay of $\rho(t)$. We tried to estimate integrated autocorrelation times $\tau_{int}$ with a self consistent truncation window of $4\tau_{int}$ [24]. The results for the integrated autocorrelation times are given in table 3. Estimates for $\tau_{int}$ are only quoted if we observed that $\rho(t)$ was positive in the entire interval from $t=0$ to $t=4\tau_{int}$. Note that $\tau_{int}$ is defined such that $\tau_{int}=0.5$ in the case of complete decorrelation. All our runs are longer than $30\,000\tau_{int}$.

In summary, all $\tau$'s are smaller or consistent with one in the range of parameters studied, with a very weak tendency to increase with increasing lattice size. Due to the ambiguities of the estimation of $\tau$ in the situation of almost complete decorrelation, we do not want to give an estimate for $z$ here. We only state that the results indicate that CSD is almost completely eliminated by the time slice blocking algorithm.

Table 2: Static observables in the two dimensional $SU(2)$ lattice gauge theory on $L \times L$ lattices, $L/\xi \approx 10$.

| $L$ | statistics | discarded | $W(\frac{L}{16}, \frac{L}{16})$ | $W(\frac{L}{8}, \frac{L}{8})$ | $\bar{P}$ | $\bar{P}^2$ |
|---|---|---|---|---|---|---|
| 16 | 100 000 | 10 000 | 0.65816(10) | 0.18751(14) | 0.0002(4) | 0.01564(7) |
| 32 | 100 000 | 10 000 | 0.67917(4) | 0.21287(12) | 0.0002(3) | 0.00855(4) |
| 64 | 50 000 | 10 000 | 0.68525(6) | 0.2204(2) | -0.0003(5) | 0.00610(5) |
| 128 | 40 000 | 5 000 | 0.68688(6) | 0.2222(3) | 0.0008(9) | 0.00545(8) |
| 256 | 40 000 | 5 000 | 0.68720(7) | 0.2232(2) | 0.0001(6) | 0.00519(4) |

Table 3: Integrated autocorrelation times $\tau_{int}$ for the two dimensional $SU(2)$ lattice gauge theory on $L \times L$ lattices, $L/\xi \approx 10$. If no value is given, we found almost complete decorrelation.

| $L$ | statistics | discarded | $\tau_{int,W(\frac{L}{16},\frac{L}{16})}$ | $\tau_{int,W(\frac{L}{8},\frac{L}{8})}$ | $\tau_{int,\bar{P}}$ | $\tau_{int,\bar{P}^2}$ |
|---|---|---|---|---|---|---|
| 16 | 100 000 | 10 000 | 0.54(1) | - | - | - |
| 32 | 100 000 | 10 000 | - | 0.60(1) | - | - |
| 64 | 50 000 | 10 000 | 0.67(1) | 0.70(1) | 0.71(1) | 0.59(1) |
| 128 | 40 000 | 5 000 | 0.76(2) | 0.74(2) | 0.92(2) | - |
| 256 | 40 000 | 5 000 | 0.88(3) | 0.83(2) | 1.01(3) | - |

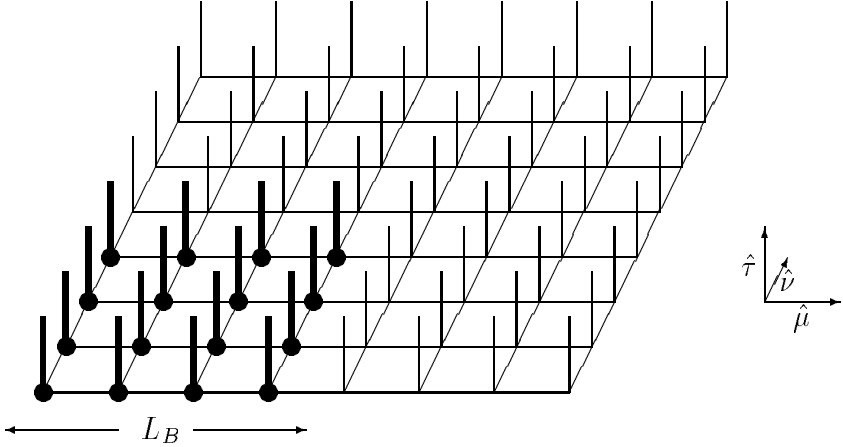

Figure 5: *Illustration of the geometry of time slice blocking in three dimensional lattice gauge theory.*

# 5 Multigrid methods for nonabelian lattice gauge fields in four dimensions

## 5.1 Covariant time slice blocking for $SU(2)$ in four dimensions

We now discuss a generalization of the covariant time slice blocking algorithm of section 3 from two dimensions to four dimensions. Many steps of the two dimensional method can be translated directly to the four dimensional case. Other features such as the nontrivial background field in the bottom of higher dimensional blocks will require a refined treatment.

Nonlocal updates can be defined as shown in figure 5 (for simplicity illustrated in three dimensions). Let us consider a fixed time direction $\tau$ with $1 \leq \tau \leq 4$ and a three dimensional time slice $\Lambda_t^\tau = \{x \in \Lambda_0 \,|\, x_\tau = t\}$. One chooses a cubic block $x'_o$ of size $L_B^3$ that is contained in $\Lambda_t^\tau$. All the link variables $U_{x,\tau}$ attached to sites $x$ inside the block $x'_o$ are proposed to be

changed simultaneously:
$$U_{x,\tau} \to U'_{x,\tau} = R_x(g) U_{x,\tau},  \qquad (5.1)$$
where $R_x(g) = g_x^* R_x g_x$ and $g_x \in SU(2)$. The rotation matrices $R_x \in SU(2)$ are again parametrized as
$$R_x(\vec{n}, s) = \cos(s\psi_x/2) + i\sin(s\psi_x/2)\,\vec{n}\cdot\vec{\sigma},  \qquad (5.2)$$
where $\vec{n}$ denotes a three-dimensional real unit vector, and the $\sigma_i$ are Pauli matrices. $\psi$ will have support on the three dimensional block $x'_o$.

Up to now the $g$-matrices are arbitrary. In the two dimensional case we have chosen them according to the block axial gauge (3.11): Choose the $g$-matrices in the one dimensional bottom of the block $x'_o$ such that $U^g_{x,\mu} = g_x U_{x,\mu} g^*_{x+\hat{\mu}} = 1$ for all $(x, x+\hat{\mu}) \in x'_o$. The gauge transformed link variables $U^g_{x,\mu}$ in the bottom of the block were equal to unity and for the case of piecewise constant interpolation the plaquettes in the interior of the block stayed unchanged (recall the discussion following figure 4). Therefore only the two plaquettes at the boundary of a block contributed to the energy change of a piecewise constant update. The energy cost was proportional to the surface of the block, not proportional to the volume of the block.

By choosing the block axial gauge in two dimensions we used the fact that we always can gauge link variables along a one dimensional open line to unity. Any nontrivial content of the gauge field in the one dimensional bottom of the block could be shifted outside of the block by applying the block axial gauge.

However in more than two dimensions, the bottom of a block will contain closed loops (see figure 5). Since a parallel transporter along a closed loop is gauge invariant, we can not get rid of the nontrivial curvature that is contained in the loop by any gauge transformation. Therefore for nontrivial gauge fields we can not find a gauge transformation $g$ such that $U^g = 1$ for all link variables in the bottom of the block $x'_o$. This means that for nontrivial gauge fields all timelike plaquettes that share a link with the bottom of the block will contribute to the average energy change of the update.

The intuitive guideline for the development of multigrid methods from section is [18]: *A piecewise constant update of a nonlocal domain should have energy costs proportional to the surface of the domain, but not energy costs proportional to the volume of the domain.*

Unfortunately, according to the discussion above, the nontrivial background field in the bottom of the block will lead to energy costs proportional to the volume of the block. Let us nevertheless attempt to have as little energy costs as possible:

Consider the extreme case of $\beta \to \infty$. Then the allowed configurations are pure gauges, i.e. configurations that are gauge equivalent to $U_{x,\mu} = 1$ for all $x, \mu$. If we choose $g$ as the transformation that brings all links to unity, it is obvious that the plaquettes in the interior of the block will not be changed by a piecewise constant update. In particular, to have this property, it is sufficient to gauge all links inside the bottom of the block to unity. This consideration leads to the following gauge condition: Choose $g$ as the gauge transformation that maximizes the functional
$$G_{C,x'_0}(U,g) = \sum_{(x,x+\hat{\mu}) \in x'_o} \mathrm{Tr}(g_x U_{x,\mu} g^*_{x+\hat{\mu}}).  \qquad (5.3)$$

We call this gauge "block Coulomb gauge". For finite $\beta$ this gauge will not bring all the links in the bottom of the block to unity, but still as close to unity as possible. Therefore the gauge field in the bottom of the block is as smooth as possible. This leads to a kind of minimization of the energy costs from the interior of the block. Note that in two dimensions the block Coulomb gauge condition reduces to the block axial gauge (3.11).

The main difference compared to the two dimensional case is that we have to expect that the energy change of the update will be proportional to the volume of the block $x'_o$. This property is caused by the fact that in four dimensions the gauge field in the bottom of the block is smooth but nonzero. This will lead to an algorithmic mass term that grows quadratic with the block dimension $L_B$. We are going to investigate the behavior of this algorithmic mass in detail below.

In summary, the steps of the nonlocal updating scheme for $SU(2)$ in four dimensions are similar to two dimensions, apart from two modifications:

1. Choose a *three dimensional* block $x'_o$ of size $L_B^3$ that is contained in the time slice $\Lambda_t^\tau$. All link variables $U_{x,\tau}$ pointing from sites $x$ inside the block in $\tau$-direction will be moved simultaneously.

2. Find the $g$-matrices defined by the *block Coulomb gauge condition*

The argument that the detailed balance condition is fulfilled by this updating scheme is analogous to the two dimensional case (cf. section 3). The only additional point is that although we only have an iterative gauge fixing algorithm in four dimensions, we do not have to fix the gauge perfectly. If we always use the same procedure in finding $g$ (e.g. a given number of relaxation sweeps starting from $g = 1$), we will always get the same $g$ and the nonlocal update is reversible.

## 5.2 Acceptance analysis for nonlocal $SU(2)$-updates

First numerical studies revealed that there is no substantial difference in the acceptance rates when instead of using the block Coulomb gauge condition one uses the Coulomb gauge condition for the whole slice $\Lambda_t^\tau$:

$$G_C(U, g) = \sum_{(x,x+\hat{\mu}) \in \Lambda_t^\tau} \text{Tr}(g_x U_{x,\mu} g^*_{x+\hat{\mu}}) \stackrel{!}{=} \text{maximal}. \tag{5.4}$$

From a practical point of view the Coulomb gauge condition is very convenient: The $g$-matrices can be calculated once and then be used for all block sizes $L_B$. In the block Coulomb gauge they would have to be recalculated for every individual block lattice. In addition, the relaxation algorithm to determine the $g$-matrices according to the Coulomb gauge condition can be vectorized in a straightforward way.

The energy change associated with the update proposal (5.1) is

$$\Delta \mathcal{H} = -\frac{\beta}{2} \sum_{\mathcal{P}} \text{Tr}(U'_{\mathcal{P}} - U_{\mathcal{P}}) = -\frac{\beta}{2} \sum_{x \in \Lambda_t^\tau} \sum_{\mu \neq \tau} \text{Tr}\{(R_x^* U_{x,\mu}^g R_{x+\hat{\mu}} - U_{x,\mu}^g) H_{x,\mu}^{g*}\}, \tag{5.5}$$

with $H^*_{x,\mu} = U_{x+\hat{\mu},\tau} U^*_{x+\hat{\tau},\mu} U^*_{x,\tau}$ and $U^g_{x,\mu} = g_x U_{x,\mu} g^*_{x+\hat{\mu}}$. $H^g$ is defined analogously. The relevant quantity for the acceptance rates is $h_1 = \langle \Delta \mathcal{H} \rangle$. For piecewise constant kernels and the gauge condition (5.4) we get

$$h_1 = 3\beta A \, (L_B - 1) L_B^2 \, \sin^2(sL_B^{-1/2}/2) + 6\beta P \, L_B^2 [1 - \cos(sL_B^{-1/2}/2)], \tag{5.6}$$

with

$$A = \langle \tfrac{1}{2}\mathrm{Tr}((U^g_{x,\mu} - \vec{n}\cdot\vec{\sigma}\, U^g_{x,\mu}\, \vec{n}\cdot\vec{\sigma}) H^{g\,*}_{x,\mu}) \rangle, \tag{5.7}$$
$$P = \langle \tfrac{1}{2}\mathrm{Tr} U_\mathcal{P} \rangle. \tag{5.8}$$

To the first term in eq. (5.6) all links contribute that are entirely inside the block, whereas the second term contains the contributions of all links that have one site in common with the surface of the block. For small $s$, the first term behaves like $s^2 L_B^2$. This is exactly the behavior of a mass term that, as we have learned in the previous sections, can be toxic for the multigrid algorithm. Note that the Coulomb gauge attempts to minimize this mass term by minimizing the quantity $A$. We identify the square root of $\beta A$ with a "disorder mass" $m_D$,

$$m_D = \sqrt{\beta A}. \tag{5.9}$$

To have a physical interpretation of $m_D$ let us discuss the "disorder scale" $l_D$ that is given by the inverse of the disorder mass:

$$l_D = \frac{1}{m_D} = \frac{1}{\sqrt{\beta A}}. \tag{5.10}$$

If the block size $L_B$ gets of the order of the disorder scale $l_D$, the mass term in eq. (5.6) becomes of order one, and the amplitudes of the nonlocal moves become suppressed. Now the crucial question is: How does the scale $l_D$ behave for large $\beta$ in comparison with the physical correlation length $\xi$? ($\xi$ is given by the inverse glueball mass or the inverse square root of the string tension.) If $l_D$ scaled with $\xi$ the algorithm could efficiently create fluctuations up to the scale of $\xi$, as required by the physics of the model. Everything would be all right if for large $\xi$

$$\frac{l_D}{\xi} \xrightarrow[\beta\to\infty]{} const. \tag{5.11}$$

In 4-dimensional $SU(2)$ lattice gauge theory the correlation length is known to increase exponentially fast with $\beta$. Therefore, for eq. (5.11) to hold, we would need that the disorder mass $m_D$ decreased exponentially fast with increasing $\beta$. Formulated differently, the crucial question is whether or not $m_D$ scales like a physical mass.

## 5.3 Monte Carlo study of $m_D$

We computed $m_D$ by Monte Carlo simulations for several values of $\beta$. For the details of these simulations see ref. [12]. There we also checked the approximation formula (2.6) for acceptance rates. It turned out to be precise also in the case of nonabelian gauge fields.

In table 5 we display the ratios of the disorder mass $m_D$ with two physical masses, the square root of the string tension $\kappa$ and the lowest glue ball mass $m_{0^+}$. The estimates for the physical masses are taken from ref. [25]. The results show that the disorder mass is nearly independent of $\beta$ in the range studied, whereas the physical masses decrease by roughly a factor of two. Thus, $m_D$ is not scaling like a physical mass for the couplings studied here. We conclude from this that for large blocks the term quadratic in $L_B$ will strongly suppress the acceptance rates even when the ratio of correlation length and block size $L_B$ is kept constant.

Table 4: Monte Carlo results for $m_D$ and $P$

| lattice size | $\beta$ | $m_D$ | $P$ | statistics |
|---|---|---|---|---|
| $8^4$ | 2.4 | 0.507(2) | 0.6305(3) | 10 000 |
| $12^4$ | 2.4 | 0.4957(4) | 0.6300(2) | 10 000 |
| $16^4$ | 2.4 | 0.4955(2) | 0.62996(5) | 10 000 |
| $8^4$ | 2.6 | 0.497(4) | 0.6703(1) | 30 000 |
| $12^4$ | 2.6 | 0.465(2) | 0.6702(1) | 20 000 |
| $16^4$ | 2.6 | 0.4644(3) | 0.67004(5) | 10 000 |
| $20^4$ | 2.6 | 0.4650(2) | 0.67008(5) | 5 000 |

Table 5: Comparison of $m_D$ with physical masses

| lattice size | $\beta$ | $m_D$ | $\sqrt{\kappa}$ | $m_{0^+}$ | $m_D/\sqrt{\kappa}$ | $m_D/m_{0^+}$ |
|---|---|---|---|---|---|---|
| $16^4$ | 2.4 | 0.4955(2) | 0.258(2) | 0.94(3) | 1.92 | 0.53 |
| $20^4$ | 2.6 | 0.4650(2) | 0.125(4) | 0.52(3) | 3.72 | 0.89 |

From this kinematical analysis it is clear that we can not expect that CSD will be eliminated by such nonlocal updates.

Let us give a plausibility argument for the large $\beta$ behavior of $m_D = \sqrt{\beta A}$: From the definition (5.7) of $A$ it is clear that $A$ vanishes for large $\beta$ because $U^g_{x,\mu}$ goes to unity in this limit. Since $A$ is a quantity that is defined on the scale of the plaquette, it is dominated by the local disorder. $A$ has nothing to do with collective excitations of the gauge field that are responsible for the formation of glueballs with a mass that decreases exponentially in $\beta$. The leading weak coupling behavior of the plaquette is [26]

$$P = 1 - \frac{3}{4\beta} + O\left(\frac{1}{\beta^2}\right) . \tag{5.12}$$

Therefore it is natural to expect a weak coupling behavior for $A$ like

$$A = \frac{c}{\beta} + O\left(\frac{1}{\beta^2}\right) , \tag{5.13}$$

with a constant $c$. If this conjecture was true we would get

$$m_D = \sqrt{\beta A} \xrightarrow[\beta \to \infty]{} const. \qquad (5.14)$$

In four dimensions the range of $\beta$ values and lattice sizes studied here is too small to check the large $\beta$ behavior of $m_D$ in detail. The analogous situation was investigated in two dimensions [18]. There, Monte Carlo estimates of $m_D$ showed the behavior $m_D \xrightarrow[\beta \to \infty]{} const.$

# 6 Multigrid Monte Carlo simulation of $SU(2)$ lattice gauge fields in four dimensions

## 6.1 Implementation of the time slice blocking

The time slice blocking scheme for $SU(2)$ in four dimensions is implemented as a recursive multigrid algorithm with piecewise constant interpolation and a W-cycle. This is feasible by updating in fixed $U(1)$ subgroups of $SU(2)$. (A detailed description of this implementation is given in [18].) This procedure is applied to a fixed time slice.

The sequence of the updating on different time slices is as follows: On an $L^4$-lattice we visit the $L$ time slices $\Lambda_t^1$, $t = 1, \ldots L$. In this way, all link variables $U_{x,1}$ on the lattice that point in the 1-direction are updated in a nonlocal way on block lattices with $L_B = 2, 4, \ldots$. Then we perform a local Creutz heat bath sweep through all links on the lattice. Now we change the time direction $\tau$ from $\tau = 1$ to $\tau = 2$: We visit the $L$ time slices $\Lambda_t^2$, $t = 1, \ldots L$, again followed by a sweep of local heat bath updates. The same scheme (a visit of all time slices and a local heat bath sweep) is repeated also for the $\tau = 3$ and $\tau = 4$ direction, such that all the link variables $U_{x,\tau}$ have been updated in a nonlocal manner for all $\tau = 1, \ldots 4$. This sequence is repeated periodically.

Observables are measured after each local heat bath sweep. In addition, we perform a random translation of the lattice after each local heat bath sweep in order to avoid effects from fixed block boundaries [16].

## 6.2 Simulation and Results

The observables measured are square Wilson loops $W(I, I) = \langle \frac{1}{2} \text{Tr}(U(C_{I,I})) \rangle$, where $U(C_{I,I})$ is the parallel transporter around a rectangular Wilson loop $C_{I,I}$ of size $I \times I$. On the $8^4$-lattice we measure $W(1,1)$, $W(2,2)$, and $W(4,4)$. Another important class of quantities is built up from timelike Polyakov loops. A Polyakov loop at the three dimensional spatial point $\vec{x}$ is defined by $P_{\vec{x}} = \frac{1}{2} \text{Tr} \prod_{t=1}^{L} U_{(\vec{x},t),4}$. We measure the lattice averaged Polyakov loop $\bar{P} = \langle 1/L^3 \sum_{\vec{x}} P_{\vec{x}} \rangle$, the lattice averaged Polyakov loop squared $\bar{P}^2 = \left\langle (1/L^3 \sum_{\vec{x}} P_{\vec{x}})^2 \right\rangle$, and the sign of the lattice averaged Polyakov loop $\text{sign}(\bar{P}) = \langle \text{sign}(1/L^3 \sum_{\vec{x}} P_{\vec{x}}) \rangle$.

In order to study the acceleration by the multigrid algorithm, we compare the autocorrelations of the nonlocal algorithm with a standard local Creutz heat bath algorithm. Precise measurements of autocorrelation times $\tau$ require high statistics simulations with run lengths of at least $1000\tau - 10\,000\tau$. For reasons of computer time we decided to simulate on relatively small lattices. The algorithms are compared on a $8^4$-lattice at $\beta = 2.2$, 2.4 and 2.6. We started the local heat bath runs from ordered configurations (all link variables set equal to unity) and discarded a suitable number of iterations for equilibration. For the multigrid simulations we used warm starts from already equilibrated configurations. Measurements were taken after each local heat bath sweep.

With these run parameters the computer time needed by our implementation on a CRAY Y-MP for one measurement on the $8^4$-lattice by the multigrid procedure is about a factor of 2.8 larger than the time needed for one measurement by the Creutz heat bath algorithm. This factor could still be lowered by using a multigrid method for gauge fixing [27].

Table 6: Comparison of the exponential autocorrelation times of heat bath (HB) and multigrid (MG) simulations for different observables on the $8^4$ lattice.

| $\beta$ | 2.2 | | 2.4 | | 2.6 | |
|---|---|---|---|---|---|---|
| algorithm | HB | MG | HB | MG | HB | MG |
| statistics | 100 000 | 50 000 | 100 000 | 100 000 | 100 000 | 100 000 |
| discarded | 10 000 | equi. | 10 000 | equi. | 10 000 | equi. |
| $\tau_{W(1,1)}$ | 6.9(2.9) | 10.3(3.5) | 5.7(1.5) | 13(9) | 1.8(6) | 1.9(1.0) |
| $\tau_{W(2,2)}$ | 6.9(1.7) | 8.0(1.8) | 26(6) | 35(7) | 4.0(1.2) | 2.9(6) |
| $\tau_{W(4,4)}$ | $\approx 1.3$ | $\approx 1.3$ | 22(3) | 26(3) | 10.4(1.8) | 10.3(2.4) |
| $\tau_{\bar{P}}$ | 3.3(5) | 3.8(6) | 93(13) | 67(4) | 279(45) | 274(33) |
| $\tau_{\bar{P}^2}$ | 1.0(3) | $\approx 1.3$ | 35(4) | 32(6) | 48(8) | 46(6) |
| $\tau_{\mathrm{sign}(\bar{P})}$ | 5.3(1.2) | 5.2(1.3) | 92(13) | 68(6) | 275(41) | 277(39) |

The results for the autocorrelation times are given in table 6. For the observable $\bar{P}$ or $\mathrm{sign}(\bar{P})$ we give a comparison of the autocorrelation functions of the multigrid algorithm with the heat bath algorithm for $\beta = 2.4$ in figure 6.

We found that the autocorrelation functions $\rho(t)$ do not in general show a pure exponential decay but exhibit a crossover from a fast mode to a slow mode that eventually governs the asymptotic decay for large $t$. Therefore the measurement of the integrated autocorrelation time $\tau_{int}$ with a self consistent truncation window of $4\tau_{int}$ might not capture the asymptotic decay of $\rho(t)$ correctly. In the present case we decided to extract a rough estimate of the exponential autocorrelation time $\tau_{exp}$ and the corresponding errors by the following procedure: We plotted the $\rho(t)$ with error bars on a logarithmic scale and decided at what value of $t$ the asymptotic exponential behavior started. Then we drew the highest and lowest straight lines that were compatible with the data in the asymptotic regime. The errors are given such that the highest and the lowest value of $\tau_{exp}$ lie at the ends of the interval result $\pm$ error.

All our results for the exponential autocorrelation times $\tau_{exp}$ of the multigrid and the heat bath algorithm are compatible within errors except from $\tau_{exp}$ for $\bar{P}$ and $\mathrm{sign}(\bar{P})$ at $\beta = 2.4$, see also figure 6. In this case the superficial gain of the multigrid algorithm is a factor of 1.4. However the computational overhead of the multigrid method compared to the heat bath algorithm (a factor of 2.8 in our implementation) was not yet taken into account. So there is no net gain in computer time also in the case of $\beta = 2.4$.

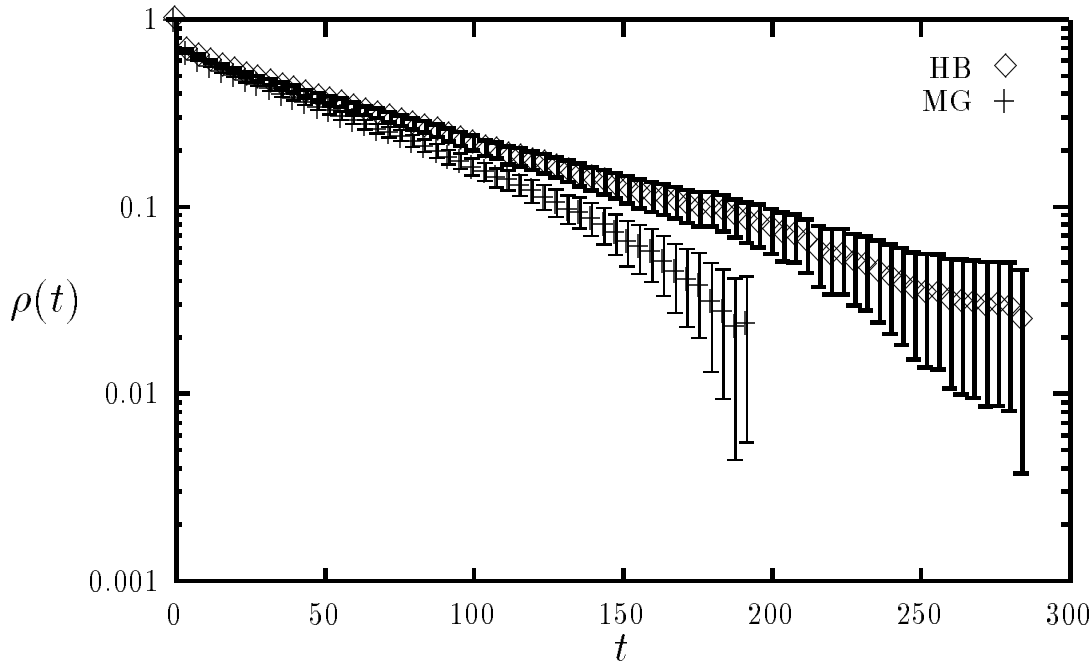

Figure 6: *Comparison of the autocorrelation functions $\rho(t)$ for the heat bath (HB) and multigrid (MG) algorithm for $SU(2)$ on the $8^4$-lattice at $\beta = 2.4$. Top: Polyakov loop $\bar{P}$, bottom: average sign of Polyakov loop $sign(\bar{P})$.*

# 7  Summary and Conclusions

We discussed multigrid algorithms for lattice gauge theory in two and four dimensions. An important observation is the statistical decoupling of adjacent time slices as long as only link variables in the time direction are updated. The time slice blocking algorithm is based on this property. The statistical independence of adjacent time slices is independent of the gauge group and of the dimensionality.

The nonabelian character of the gauge field was first discussed in two dimensions. We proposed the gauge covariant time slice blocking for $SU(2)$ in two dimensions that is particularly adapted to the nonabelian character. From the kinematical analysis we could expect a strong reduction of CSD in a simulation of the proposed algorithm.

A multigrid Monte Carlo algorithm using the time slice blocking method with piecewise linear interpolation was implemented in the unigrid style with a V-cycle. We used a heat bath version in randomly chosen $U(1)$ subgroups. The simulations were performed on physically large lattices of the size $10\xi \times 10\xi$.

All observed integrated autocorrelation times were found to be smaller or consistent with one on lattice sizes up to $256^2$, with a very weak tendency to increase with increasing lattice size. Therefore we conclude that the time slice blocking algorithm eliminates CSD almost completely.

It is fair to say that we used a special feature of two dimensional lattice gauge theory: In the infinite volume limit or for open boundary conditions one can decouple two dimensional lattice gauge theory to a set of independent one dimensional spin models by choosing the axial gauge. Although we use periodic boundary conditions here, our method is very much in the spirit of updating on independent time slices.

We discussed whether this concept is still successful if we generalize it to four dimensions. There, new difficulties occur that are due to the nontrivial background field in the bottom of the blocks. We investigated the scale dependence of acceptance rates in detail. Here we found that an algorithmic mass term generated by the disorder in the background field suppresses the acceptance rates on large blocks. From our kinematical analysis we can not expect that the proposed algorithm will have a chance to reduce the dynamical critical exponent below $z \approx 2$. However, compared to local Monte Carlo algorithms one could still find an acceleration of the dynamics by a constant factor, depending on the details of the implementation.

This question was adressed by an implementation of the multigrid Monte Carlo algorithm in four dimensions. We used the time slice blocking method and piecewise constant interpolation with a W-cycle in a recursive multigrid version by updating in $U(1)$ subgroups of $SU(2)$. Simulations with gauge couplings $\beta = 2.2$, 2.4 and 2.6 were performed on an $8^4$-lattice.

Apart from a modest acceleration of Polyakov loop observables at $\beta = 2.4$ by a factor of 1.4, no improvement was found compared to a local heat bath algorithm. Since the nonlocal update procedure has a computational overhead of a factor of 2.8 on the $8^4$-lattice (on a CRAY Y-MP), there is no net gain but even a loss in CPU-time. This factor depends however on the details of the implementation.

Since from the theoretical analysis (section 5) one can not expect a lower dynamical critical exponent $z$, there is no hope that on larger lattices the method will perform better.

Possible improvements of our nonlocal updating scheme were investigated by Gutbrod [28]. Starting from the Coulomb gauge, he uses an additional smoothing by taking a quadratic approximation to the action and updating in terms of approximate eigenfunctions. The resulting nonlocal updates are performed as (approximately) microcanonical overrelaxation steps. Asymmetric lattices of size $L^3 T$ with $T \gg L$ and anisotropic couplings (e.g. $\beta = 4.5$ in the time direction and $\beta = 3.0$ in the space directions) are used. The results indicate a superficial gain of a factor of $1.5 - 3$ compared to a local overrelaxation algorithm. However if the (implementation dependent) computational overhead of the nonlocal method is taken into account, the net gain is marginal.

# Acknowledgments

We would like to thank F. Gutbrod, G. Mack and A. D. Sokal for helpful discussions. This work was supported in parts by the Deutsche Forschungsgemeinschaft, the German Israeli Foundation and the MINERVA Foundation. The simulations were performed on the CRAY Y-MP of the HLRZ in Jülich.